\documentclass[12pt]{dcds2}

\def\ppages{1--10}

\newcommand{\be}{\begin{equation}}
\newcommand{\ee}{\end{equation}}
\newcommand{\bea}{\begin{eqnarray}}
\newcommand{\eea}{\end{eqnarray}}

\theoremstyle{plain}

\theoremstyle{definition}

\title[]
{Quantum Chaos via the Quantum Action}

\email{hkroger@phy.ulaval.ca}

\author[Helmut Kr\"{o}ger]{}

\begin{document}
\maketitle

\centerline{\scshape Helmut Kr\"{o}ger}

\medskip

{\footnotesize \centerline{} 
\centerline{D\'{e}partement de Physique}
\centerline{Universit\'{e} Laval}
\centerline{Qu\'{e}bec, Qu\'{e}bec G1K 7P4, Canada}
 }

\begin{quote}{\normalfont\fontsize{8}{10}\selectfont
{\bfseries Abstract.}
We discuss the concept of the quantum action with the purpose to characterize and quantitatively compute quantum chaos. As an example we consider in
quantum mechanics a 2-D Hamiltonian system  - harmonic oscillators with anharmonic coupling -  which is classically a chaotic system. We
compare Poincar\'e sections obtained from the quantum action with those from the classical action. 
\par}
\end{quote}

\section{Introduction}
\label{sec:Intro} 

The phenomenon of chaos is quite well understood in classical physics. 
Its origin in terms of nonlinear dynamics has been investigated thoroughly 
in theoretical physics and mathematics. 
If one considers nature at the level of atoms, where the laws of quantum mechanics (Q.M.) hold, chaotic phenomena are by far less well understood than 
in classical (macroscopic) physics \cite{Blumel,Gutzwiller,Haake,Nakamura,Stockmann}.
Many questions are unresolved yet. For example one can ask the following questions:
What is characteristic for quantum chaos? Which are its essential features?
What is the analogue of classical Lyapunov exponents in Q.M.?
What is the analogue of classical Poincar\'e sections and of classical KAM surfaces in Q.M.?  

Here we take the point of view that the quantum action \cite{Jirari01a,Jirari01b,Jirari01c,Caron01,Kroger02} is the ideal tool to 
address those questions. In the following we discuss the concept of the quantum action. In Sect.\ref{sec:QuantumAction} we discuss its physical meaning, 
the question of existence in the mathematical sense, its symmetry properties
and methods of its analytical and numerical computation. In Sect.\ref{sec:Analytical} we present analytical results for the quantum action in the limit when temperature goes to zero.
In Sect.\ref{sec:QuantChaos} we discuss quantum chaos at hand of the 2-D anharmonic oscillator.
Finally, we give an outlook on further use of the quantum action, e.g.
in cosmology, in the inflationary scenario, 
in condensed matter, and in particular in supraconductivity, 
and in quantum chaos with time-dependent Hamiltonians.

\section{Quantum action}
\label{sec:QuantumAction}

We start out by asking: What is problem of chaos physics when making the transition from classical to quantum mechanics?
Let us recall classical mechanics: Equations of motions
are given by specifying the dynamics (usually in terms of a differential equation) and by specifying initial conditions. Usually one specifies position and velocity at some initial time. 
A consequence of this type of boundary conditions is that certain trajectories are forbidden. For example those where the energy is too small to overcome a potential barrier.
This type of boundary conditions has more severe consequences when going over to quantum physics: No Q.M. transition is compatible with such initial conditiuon because Heisenberg's uncertainty relation prohibits to specify both, position and momentum with zero error. This is the cause why Lyapunov exponents can not be carried over from classical physics to quantum physics. This is a severe 
problem for quantum chaos.
However, there is a simple way out: Reconsider classical mechanics with 2-point boundary conditions. I.e., one specifies positions at some initial and final time. 
Now the consequence is: Classically, there are no forbidden trajectories. (Sometimes even infinitely many trajectories are possible!)
In quantum physics, there is a well defined transition amplitude.

\subsection{Motivation of quantum action}
\label{sec:Motivation} 

The physical idea of the quantum action has a threefold motivation.
First, it can be viewed as a result of renormalisation \cite{Jirari01b}.
Renormalisation is well known from solid state physics. When a charged particle propagates in a solid it interacts with atoms. As a result, the properties
of the propagating charged particle can be described by an effective mass and charge different from that of free propagation, i.e. mass and charge are renormalized. Similar mechanism are well known to occur in quantum field theory (QED, QCD).
Here we suggest that the difference in propagation between 
Q.M. and classical physics (Q.M. fluctuations, zig-zag paths with Hausdorff dimension $d_{H}=2$) can be viewed as a renormalisation of the classical action to become the so-called quantum action. Q.M. fluctuations are manifest in mass and potential parameters different from the classical ones.  

Second, as a guide to construct such renormalized action
one can use Feynman's path integral.
In special cases Feynman's path integral becomes very simple, namely it becomes a sum over classical pathes only \cite{Schulman}
\begin{equation}
\label{SumClassPath}
G(x_{f},t_{f}; x_{i},t_{i}) = 
\sum_{ \{x_{cl}\} } Z \exp \left[ \frac{i}{\hbar} 
\left. S[{x}_{cl}] \right|_{x_{i},t_{i}}^{x_{f},t_{f}} \right] ~ .
\end{equation}
The most well known example is the harmonic oscillator given by the potential $V(x) = \frac{m}{2} \omega^{2} x^{2}$,
\begin{eqnarray}
\left. S[x_{cl}] \right|_{x_{i},t_{i}}^{x_{f},t_{f}} 
&=& \frac{ m \omega}{2 \sin(\omega T) } 
\left[ (x_{f}^{2} + x_{i}^{2}) \cos(\omega T) - 2 x_{i}x_{f} \right] , 
\nonumber \\
Z &=&  \sqrt{ \frac { m \omega }{ 2 \pi i \hbar \sin(\omega T) } }, ~~~
T = t_{f}-t_{i} . 
\end{eqnarray}
However, such simple relation is not true in general! 
One expects the quantum action to be different from the classical action, in general. 

Third, we shall ask:  What is the mathematical form of the quantum action? Some hint comes from symmetry. For a given positive number $\alpha$ consider the following scale transformation involving the mass $m$, the potential $V(x)$ and the transition time (time elapsed during evolution of the system) $T$, 
\begin{eqnarray}
m &\to& m / \alpha
\nonumber \\
V(x) &\to& \alpha V(x)
\nonumber \\
T &\to& T / \alpha ~ .
\end{eqnarray}
This transforms the quantum mechanical Hamiltonian 
and quantum mechanical transition amplitude as follows:
\begin{eqnarray*}
H &\to& \alpha H
\nonumber \\
<b| e^{-i\hat{H}T/\hbar} |a> &\to& \mbox{invariant} ~ .
\end{eqnarray*}
On the other hand, this symmetry transformation has the following implications in classical physics:
It transforms the classical trajectory (solution of Euler-Lagrange equation of motion), the Lagrangian and the action evaluated along the classical trajectory in the following way: 
\begin{eqnarray*}
x_{cl}(t) &\to& x_{cl}(\alpha t) 
\nonumber \\
L(x_{cl}(t),\dot{x}_{cl}(t)) &\to& 
\alpha L(x_{cl}(\alpha t),\dot{x}_{cl}(\alpha t)) 
\nonumber \\
S[x_{cl}] &\to& \mbox{invariant} ~ .
\end{eqnarray*}
As a result we find that the Q.M. transition amplitude and any local classical action evaluated at its classical trajectory are all invariants. 
Note that the Lagrangian underlying the quantum mechanical Hamiltonian and the Lagrangian of the classical system need {\it not} be the same. I.e. this symmetry quite genral (but requires a local potential).
We interpret this observation as a hint that the quantum transition amplitude may have something to do with some action evaluated along its classical trajectory. As consequence we formulate the following postulate of the quantum action.

\subsection{Postulate of quantum action}
\label{sec:Postulate}  

The quantum action \cite{Jirari01a} is defined by the following requirements.
For a given classical action 
\begin{equation}
S[x] = \int dt \frac{m}{2} \dot{x}^{2} - V(x) ~ ,
\end{equation}
there is a quantum action
\begin{equation}
\tilde{S}[x] = \int dt \frac{\tilde{m}}{2} \dot{x}^{2} - \tilde{V}(x) ~ ,
\end{equation}
which parametrizes the Q.M. transition amplitude 
\begin{eqnarray*}
&& G(x_{f},t_{f}; x_{i},t_{i}) = \tilde{Z} 
\exp \left[ \frac{i}{\hbar} 
\left. \tilde{\Sigma} \right|_{x_{i},t_{i}}^{x_{f},t_{f}} \right] ~ , 
\nonumber \\
&& \left. \tilde{\Sigma} \right|_{x_{i},t_{i}}^{x_{f},t_{f}} 
= \left. \tilde{S}[\tilde{x}_{cl}] \right|_{x_{i},t_{i}}^{x_{f},t_{f}}  
= \left. \int_{t_{i}}^{t_{f}} dt ~ \frac{\tilde{m}}{2} \dot{\tilde{x}}_{cl}^{2} - \tilde{V}(\tilde{x}_{cl}) \right|_{x_{i}}^{x_{f}}  ~ .
\end{eqnarray*}
where $\tilde{x}_{cl}$ denotes the classical path corresponding to the action $\tilde{S}$. We require 2-point boundary conditions,
\begin{eqnarray*}
&& \tilde{x}_{cl}(t=t_{i}) = x_{i}
\nonumber \\
&& \tilde{x}_{cl}(t=t_{f}) = x_{f} ~ .
\end{eqnarray*}
$\tilde{Z}$ stands for a dimensionful normalisation factor. 
Note: The parameters of the quantum action (mass, potential) are independent of the boundary points $x_{f}$, $x_{i}$, but depend on the transition time  
$T=t_{f}-t_{i}$. The same q-action parametrizes all transition amplitudes for a given transition time. 

The postulate leads to the question: Does such quantum action exist?
So far we have no general proof. However, the answer is affirmative in the following cases:
\newline (a) Example: Harmonic oscillator. In this case the classical action satisfies the definition of the quantum action \cite{Schulman,Jirari01a}, hence both coincide.
\newline (b) Arbitrary local potential, limit when transition time $T \to 0$: 
Then again the quantum action exists (coincides with classical action). 
\newline (c) Arbitrary potential, imaginary time, limit when transition time  
$T \to \infty$ (equivalent to temperature going to zero): 
Then the quantum actions exists, being different from the classical action, in general \cite{Kroger02}.

\section{How to compute the quantum action}
\label{sec:Compute} 

So far we have explored two methods to determine the quantum action: (A) Global optimisation of parameters of quantum action to fit quantum mechanical transition amplitudes \cite{Jirari01a,Jirari01c,Caron01}. According to the definition of the quantum action (sect.\ref{sec:Postulate}), a quantum action with one set o parameters should 
well represent Q.M. transition amplitudes for all combinations of initial and final boundary points. Thus we compute a large but finite set of transition amplitudes (via solution of the Schr\"odinger equation, eigen function expansion methods, or solution of the path integral via Monte Carlo) and adjust the unkonwn parameters of the quantum action (mass, coefficients of the potential) until a globally satisfactory fit is achieved \cite{Jirari01a}.

An alternative method (B), a renormalisation group flow equation has been proposed in Ref.\cite{Jirari01b}. There is some analogy with quantum field theory (QFT). In QFT action parameters depend on scale parameters (cut-off $\Lambda$, lattice spacing $a_{s}$ and $a_{t}$).
The dependence of the action on the the scale parameters is ruled by renormalisation group equations (e.g. Callan-Symanzik).
In our case we consider quantum mechanics. Usually it is not difficult to enter the regime close to the continuum limit (when $\Delta s$ and $\Delta t$  
become small).
The flow equation aims to describe the flow of the parameters of the quantum action under variation of the parameter transition time $T$ (or inverse temperature $\beta$). This equation is obtained by considering the Schr\"odinger equation for the Q.M. transition amplitude and inserting the ansatz of the quantum action. The resulting flow equation is given in Ref.\cite{Jirari01b}.
The flow equation method has the advantage that it does not require to compute
explicitely the Q.M. transition amplitudes, which is computationally costly.

\section{Analytical properties of the quantum action} 
\label{sec:Analytical}

For physical and mathematical reasons, it is interesting to go from real time over to imaganiary time. A mathematical reason is that the path integral for the Q.M. transition amplitude then becomes well defined (Wiener integral). A physical reason is that the description of finite temperature physics requires the use of imaginary time. Thus we make the following transition
\begin{equation*}
t \to -it ~ .
\end{equation*}
Then the action becomes the Euclidean action 
\begin{equation}
S_{E} = \int_{0}^{T} dt \frac{m}{2} \dot{x}^{2} + V(x) ~ ,
\end{equation}
the transition amplitude becomes the Euclidean transition amplitude
\begin{eqnarray*}
G_{E}(x_{f},T;x_{i},0) &=& \langle x_{f} | \exp[ - H T/\hbar ] | x_{i} \rangle
\nonumber \\
&=& \left. \int [dx] \exp \left[ - S_{E}[x]/\hbar \right] \right|_{x_{i},0}^{x_{f},T} ~ ,
\end{eqnarray*}
and the quantum action becomes the Euclidean quantum action 
\begin{equation}
G_{E}(x_{f},T; x_{i},0) = \tilde{Z}_{E} 
\exp \left[ -\frac{1}{\hbar} 
\left. \tilde{\Sigma}_{E} \right|_{x_{i},0}^{x_{f},T} \right] ~ . 
\end{equation}

Before we present some analytic results, let us recall the Feynman-Kac limit:
The time evolution of a quantum system in imaginary time is determined by 
the behavior of $\exp[- H T/\hbar]$ when $T$ goes to $\infty$. In the Feynman-Kac limit one has
\begin{equation*}
e^{- H T/\hbar } 
\longrightarrow_{T \to \infty} 
|\psi_{gr} \rangle e^{- E_{gr} T/\hbar} \langle \psi_{gr} | ~ ,
\end{equation*}
where $E_{gr}$ and $\psi_{gr}$ are the ground state energy and wave function, respectively.

Now we consider the Euclidean action in the limit of large time. 
The trajectory minimizes the action. Because the kinitic term and the potential term (by assumption) are positive, this means that the trajectory minimizes fluctuations and stays as close as possible to the bottom of the potential valley. This implies for $T \to \infty$,
\begin{equation*}
\tilde{V} \to \tilde{V}_{min} ~ , ~~~ 
\tilde{T}_{kin} \to 0 ~ , ~~~
\epsilon = - \tilde{T}_{kin} + \tilde{V} \to \tilde{V}_{min} ~ ,
\end{equation*}
and
\begin{equation*}
\tilde{\Sigma} \equiv \tilde{S}[\tilde{x}_{cl}]|_{x_{i},0}^{x_{f},T} 
= \int_{0}^{T} dt ~ \tilde{T}_{kin} + \tilde{V} 
= \tilde{v}_{0} T + \left( 
\int_{x_{i}}^{0} + \int_{0}^{x_{f}} dx ~  
_{\pm} \sqrt{2 \tilde{m}(\tilde{V}(x) -\tilde{V}_{min}) } 
\right) ~ ,
\end{equation*}
where the sign depends on initial and final data. 
The transition amplitude then becomes
\begin{eqnarray*}
&&G(x_{f},T;x_{i},0) = \left. \tilde{Z} \exp[- \tilde{\Sigma}/\hbar] \right|_{x_{i},0}^{x_{f},T} 
\stackrel{T \to \infty}{\longrightarrow} 
\tilde{Z}_{0} ~ \exp[ -\tilde{v}_{0} T/\hbar ]
\nonumber \\
&&\times
\exp[ - \int_{0}^{x_{fi}} dx ~ 
\sqrt{2 \tilde{m}( \tilde{V}(x) -\tilde{V}_{min} ) }/\hbar  ]
\exp[ - \int_{x_{in}}^{0} dx ~ 
\sqrt{2 \tilde{m}( \tilde{V}(x) - \tilde{V}_{min} ) }/\hbar ] ~ . 
\end{eqnarray*}
By comparison with the Feynman-Kac formula one obtains
the following analytic expressions for the ground state energy and wave function, expressed in terms of the quantum action,
\begin{equation}
\label{eq:QPotExtr}
E_{gr} = \tilde{V}_{min} ~ , ~~~ 
\psi_{gr}(x) = \frac{1}{N} ~ e^{ - \int_{0}^{|x|} dx' ~ 
\sqrt{2 \tilde{m}( \tilde{V}(x') - \tilde{V}_{min} ) }/\hbar } ~ .
\end{equation}
Combining this with the Schr\"odinger equation leads to the following transformation law
\begin{equation}
\label{eq:TransLaw}
2 m(V(x) - E_{gr})  
=2 \tilde{m}(\tilde{V}(x) - \tilde{V}_{min}) 
- \frac{\hbar}{2} \frac{ \frac{d}{dx} 2 \tilde{m} (\tilde{V}(x) - \tilde{V}_{min})}
{ \sqrt{2 \tilde{m}( \tilde{V}(x) - \tilde{V}_{min} ) } } ~ \mbox{sgn}(x) ~ .
\end{equation}
Note: Although those results have been obtained in imaginary time,
they hold also in real time (ground state energy $E_{gr}$ and 
wave function $\psi_{gr}(x)$ are the same as in real time).
Remarkable properties are the following: Eq.(\ref{eq:QPotExtr}) shows that 
the minimum of the quantum potential coincides with the ground state energy.
Second, the position of the minimum of the quantum potential coincides with the position of the maximum of the ground state wave function. This shows that the quantum potential gives a much better picture of the behavior of the quantum system than the classical potential, which does not all share those properties.

\subsection{Quantum action and WKB formula} 
\label{sec:WKB}

The celebrated semi-classical WKB formula [Wentzel, Kramers and Brillouin (1926)] gives approximate solutions to wave function and tunneling amplitude. 
As an example let us consider in 1-D a system with parity symmetric "confinement" potential (potential goes to $+ \infty$ when $|x| \to \infty$). 
Assume that the system is in the ground state. 
The WKB formula gives an approximate expression for wave function at energy $E=E_{gr}$
\begin{equation*}
\psi_{WKB}(x) = \frac{A}{(2m[V(x) - E_{gr}])^{1/4}} ~
\exp\left[ - \frac{1}{\hbar} \int_{x_{0}}^{x} dx' \sqrt{2m[V(x') - E_{gr}]} \right] 
~ .
\end{equation*}
Comparing with the results from sect.\ref{sec:Analytical},
we see that the WKB formula becomes exact, when replacing 
parameters of classical action by those of the quantum action
\begin{equation*}
m \to \tilde{m} ~ , ~~~ 
V(x) \to \tilde{V}(x) ~ , 
\end{equation*}
and replacing the rational term in front by a constant (wave function normalisation).

\section{Excited states of hydrogen atom} 
\label{sec:Hydrogen}

Can the quantum action also give analytical results for excited states?
The answer is yes, if we consider excited states being lowest energy states of a conserved quantum number.
As an example let us consider the radial motion of the hydrogen atom in a sector of fixed angular momentum $l>0$. 
The potential has a centrifugal plus a Coulomb term. 
Let us consider angular motion to be quantized.
We keep the angular momentum quantum number $l$ fixed. 
In the Feynman-Kac limit the transition amplitude is projected onto the state of lowest energy compatible with quantum number $l$. 
The states of the hydrogen atom are characterized 
by the quantum numbers $n$ (principal quantum number) $l$ (angular momentum) where $n=l+1$. There is also the magnetic quantum number $m$.
The radial potential is given by
\begin{equation*}
V_{l}(r) = \frac{\hbar^{2}l(l+1)}{2m r^{2}} - \frac{e^{2}}{r} ~ .
\end{equation*}
The exact energy of the excited states is given by
\begin{equation*}
E_{l} = - \frac{E_{I}}{n^{2}}= - \frac{E_{I}}{(l+1)^{2}} ~ , ~~~ 
E_{I} = \frac{m e^{4}}{2 \hbar^{2}} \approx 13.6 eV ~ \mbox{(ionisation energy)} ~ .
\end{equation*} 
The corresponding wave function is given by
\begin{equation*}
\phi_{l}(r) = \frac{1}{N_{l}} ~ \left(\frac{r}{a_{0}}\right)^{l} ~ 
\exp\left[ -\frac{r}{(l+1) a_{0}} \right] ~ , ~~~
a_{0} = \frac{\hbar^{2}}{m e^{2}} ~ \mbox{(Bohr radius)} ~ . 
\end{equation*}
For the quantum action we make an ansatz 
\begin{equation*}
\tilde{m}=m ~ , ~~~
\tilde{V}_{l}(r) = \mu/r^{2} -\nu/r ~ .
\end{equation*}
A transformation law [similar to Eq.(\ref{eq:TransLaw})] determines the parameters of the q-action,
\begin{equation*}
\mu = \frac{\hbar^{2}}{2m} l^{2} ~ , ~~~
\nu = e^{2} \frac{l}{l+1} ~ ,
\end{equation*}
and the minimum of the quantum potential gives exactly the excitation energies
\begin{equation*}
E_{l} = \tilde{V}^{min}_{l} = - \frac{m e^{4}}{2 \hbar^{2}} \frac{1}{(l+1)^{2}} = - \frac{E_{I}}{(l+1)^{2}} ~ .
\end{equation*}
Again the wave function can be expressed in terms of the quantum action in a way similar to Eq.(\ref{eq:QPotExtr}), and reproduces the exact wave function $\phi_{l}$.
Moreover, we observe that the excited state wave function $\phi_{l}(r)$ has its maximum where the quantum potential $\tilde{V}_{l}$ has its minimum.
The quantum action has the same structure as the classical action. 
Both, the centrifugal and the Coulomb term get tuned.

\section{Quantum chaos}
\label{sec:QuantChaos}
The hydrogen atom in the presence of strong magnetic fields has been explored experimentally and theoretically \cite{Bohigas,Friedrich, Wintgen}.
It displays quantum chaos via disorder in the spectrum. In the regime of high lying excited states, where the system becomes semi-classical, Gutzwiller's trace formula has been applied successfully by Wintgen \cite{Wintgen} to extract periodic orbit information. Here we suggest to define and explore quantum chaos 
via the phase space generated by the quantum action. As a prototype system, we consider a 2-dim non-integrable Hamiltonian, with a classically chaotic counter part. We employ the following definition of chaos in a quantum system: \\
 
\noindent {\bf Definition of quantum chaos:} \\
{\em Consider a classical system with action $S$. 
The corresponding quantum system 
displays quantum chaos, if the corresponding quantum action $\tilde{S}$ in the asymptotic regime $T \to \infty$ generates a chaotic phase space.}  \\

Some comments are in order. First, our definition of quantum chaos is based on the concept of 'some' phase space related to quantum mechanics. The novel idea is to introduce this phase space via the quantum action. Second, the quantum action depends on the transition time $T$, however, for large $T$ it converges  
asymptotically. Moreover, in this regime the existence of the quantum action has been established rigorously. The regime of large $T$ makes physical sense, because the proper definition of Lyapunov exponents, one of the characteristics of chaotic dynamics, involves the large time limit. 

As an example, let us consider the following Hamiltonian
\begin{eqnarray*}
&& S = \int_{0}^{T} dt ~ \frac{1}{2} m (\dot{x}^{2} + \dot{y}^{2}) 
+ V(x,y) 
\nonumber \\
&& V(x,y) = v_{2}(x^{2} + y^{2}) + v_{22} x^{2}y^{2} 
\nonumber \\
&& m = 1 ~ , ~~~ 
v_{2} = 0.5 ~ , ~~~
v_{22} = 0.05 ~ .
\end{eqnarray*}
The parameter $v_{22}$ controls the deviation from integrability ($v_{22}=0$).
Because in the case $v_{22}=0$ the system corresponds to two uncoupled oscillators, the corresponding quantum action then coincides with the classical action. 

For the quantum action, we have made the following ansatz: It reflects the time-reversal symmetry, parity conservation and symmetry under exchange 
$x \leftrightarrow y$:
\begin{eqnarray*}
\tilde{S} &=& \int_{0}^{T} dt ~ 
\frac{1}{2} \tilde{m} (\dot{x}^{2} + \dot{y}^{2}) 
+ \tilde{V}(x,y), ~~~
\nonumber \\
\tilde{V} &=& \tilde{v}_{0} 
+ \tilde{v}_{2} (x^{2} + y^{2}) 
+ \tilde{v}_{22} x^{2}y^{2} 
+ \tilde{v}_{4} (x^{4} + y^{4}) ~ .
\end{eqnarray*}

We also have included in the quantum action terms like 
$\dot{x} \dot{y}$,
$xy$,
$xy^{3}+x^{3}y$,
$x^{2}y^{4} + x^{4}y^{2}$,
$x^{4}y^{4}$. 
Numerically, those coefficients were found to be small (compared to machine precision or zero within error bars).

Numerical studies \cite{Caron01} have shown the following behavior:
For small $v_{22}$, Poincar\'e sections of classical and quantum physics are quite similar. With increase of energy, both display mixed dynamics and become more chaotic. Also, with increase of energy differences between classical and quantum phasse space become more pronounced. Islands of regular behavior differ in shape and position. Quantitatively, one observes that the value of 
$\tilde{v}_{22}$ (quantum action) is smaller than the corresponding value 
$v_{22}$ of the classsical action. Because this parametyers measures the deviation from integrability, this hints to the possibility that the quantum system is "less" chaotic than the classical system. A more detailed analysis is called for.

\section{Outlook: Further use of quantum action.}
\label{sec:Outlook}

In cosmology, in particular in the inflationary scenario, it is important for the evolution of the universe (galaxy formation etc.) to search for new minima of the quantum potential and corresponding quantum instantons. The quantum action is suitable to describe quantum instantons \cite{Jirari01c}.

In elementary particle physics, instantons play a role in 
gauge field theories. They are believed to play a role in the 
mecanism of quark confinement, for the description of baryonic matter at high temperature and density and in the phase transition to quark-gluon plasma (early universe, neutron stars). New phases (superconducting?) have been predicted (Princeton group) where instanton effects are important. The quantum action may help to find quantum instantons.

In condensed matter physics and 
supraconductivity, for the study of quantum chaos in Josephson junctions Gutzwiller's trace formula has been applied. The quantum action may be useful to treat quantum chaos in periodic potentials.   

Finally, in the context of quantum chaos in atomic physics, chaos assisted tunnelling between islands of stability has been recently discussed as a hot topic. Phase space has been treated classically in such studies. The quantum action may be useful to draw a better portrait of quantum phase space. \\

\noindent {\bf Acknowledgements} \\ 
H.K. is grateful for support by NSERC Canada.

\end{document}